\documentclass[graybox]{svmult}

\usepackage{type1cm}        
\usepackage{makeidx}         
\usepackage{graphicx}        
\usepackage{multicol}        
\usepackage[bottom]{footmisc}

\usepackage{amsmath,mathtools}
\usepackage{newtxtext}       %
\usepackage[varvw]{newtxmath}

\usepackage[square,comma,numbers,compress]{natbib}
\colorlet{mylinkcolor}{blue!66!black!80}
\usepackage[colorlinks=true,linkcolor=mylinkcolor,citecolor=mylinkcolor,urlcolor=mylinkcolor]{hyperref}%
\usepackage{color}

\newcommand{\e}{{\rm e}}
\newcommand{\mtau}{\langle\tau\rangle}
\newcommand{\ntau}{\overline{\tau}_n}

\newcommand{\C}{\mathcal{C}}

\newcommand{\blue}[1]{\textcolor{black}{#1}}

\makeatletter
\newcommand\thefontsize{The current font size is: \f@size pt}
\makeatother

\makeindex             

\begin{document}

\title*{Concentration of Empirical First-Passage Times}
\author{Rick Bebon and Alja\v{z} Godec}
\institute{Rick Bebon \at Mathematical bioPhysics Group, Max-Planck-Institute for Multidisciplinary Sciences, 37077 Göttingen, Germany, \email{rick.bebon@mpinat.mpg.de}
\and Alja\v{z} Godec \at Mathematical bioPhysics Group, Max-Planck-Institute for Multidisciplinary Sciences, 37077 Göttingen, Germany \email{agodec@mpinat.mpg.de}}
%
%
\maketitle
\abstract*{}
\abstract{
First-passage properties 
are central to the kinetics 
of target-search processes. Theoretical approaches so far primarily focused on predicting first-passage statistics
for a given process or model. In practice, however, one faces the
reverse problem of inferring first-passage statistics from, typically sub-sampled,
experimental or simulation data.
Obtaining trustworthy estimates
from under-sampled data and unknown underlying dynamics remains a
daunting task, and the assessment 
of the uncertainty 
is imperative.
In this chapter, we highlight 
recent progress in understanding and controlling finite-sample effects
in empirical first-passage times of reversible Markov processes.
Precisely, we present concentration inequalities bounding from above
the deviations of the sample mean for any sample size from the true mean
first-passage time and construct \emph{non-asymptotic} confidence
intervals. Moreover, we present two-sided bounds on the range of
fluctuations, i.e, deviations of the expected maximum and minimum from the mean in any
given sample, which control uncertainty even in situations where
the mean is \emph{a priori} not a sufficient statistic.
}

\section{Introduction}
In the past decades, there has been a surge of interest in 
target-search problems, typically devoted to determining the
time it takes a  
random searcher to find a 
target.
This stochastic time, canonically referred to as \emph{first-passage}
or \emph{first-hitting time},
has proven to be invaluable in understanding  
kinetic properties of various physical, chemical, and
biological processes~\cite{redner2001guide, metzler2014first, lindenberg2019, iyer2016first, Zhang2016}.

As a pioneering example consider diffusive barrier crossing, when
a thermally driven searcher (e.g., a reactant molecule) must overcome
an energy barrier  
in order to ``collide'' with its reaction or binding partner.
This situation arises naturally in the classical picture of Kramers' reaction-rate 
theory~\cite{Kramers,Haenggi},
according to which kinetic rates of chemical reactions
can be understood in terms of inverse mean first-crossing times. 
Building on the success of these ideas, first-passage theory remained an effective paradigm for investigating (bio-) chemical 
kinetics ever since~\cite{Szabo,Redner,Grebenkov_2018,Grebenkov_2018_2,Grebenkov_PRL1,Bressloff,Roldn2016,
Ghusinga2017,Rijal2022,Parmar2015,Charlebois2011,Frey2019,Lloyd2001,Hufnagel2004,Olivier_RMP,Boccardo2022,
Erdmann2004_2,Thirumalai,Blom2021,Goychuk2002}.

In a similar vein, numerous events in biological settings are initiated 
when a diffusing searcher, for example, a messenger ligand, first binds to 
its target, its receptor.
Evidently, this prompts the application of first-passage ideas to study
the kinetics of molecular and cellular processes such as
cell signaling and gene regulation~\cite{Berg,Andy,Holcman,Olivier_Chromatin,Marklund,Bauer,Olivier_NC,Bnichou2014,Godec_PRX,Newby_PRL},
intracellular transport~\cite{Bressloff},
RNA biosynthesis~\cite{Roldn2016},
stochastic protein accumulation~\cite{Ghusinga2017,Rijal2022},
DNA-binding~\cite{Parmar2015},
virus uptake~\cite{Frey2019},
cell adhesion~\cite{Erdmann2004_2,Thirumalai,Blom2021},
kinetochore capture~\cite{Nayak2020},
or gating of ion channels~\cite{Goychuk2002}.
Notably, going beyond the molecular and cellular scope, first-passage principles 
have also found biological relevance in exploring
the emergence of drug resistance~\cite{Charlebois2011},
spreading dynamics of diseases~\cite{Lloyd2001,Hufnagel2004},
and foraging behavior of bacteria and animals~\cite{Olivier_RMP}.

It should thus come to no surprise that addressing 
target-search questions through the lens of first-passage 
properties led to an immense body of literature 
with successful applications spanning 
virtually all scientific domains, including 
mathematics, physics, chemistry, biology, geology, economics, and finance, among others 
(see e.g.~\cite{redner2001guide,metzler2014first,lindenberg2019, schuss2009theory,bouchaud2003theory,Paul2013,Gardiner}
and references therein).  

First-passage concepts have further 
been applied in more abstract settings
to characterize persistence 
properties~\cite{Dougherty2002,Constantin2003,Merikoski2003,Constantin2004,Dougherty2005,Godrche2009,Bray_2013},
as well as diffusion through interfaces~\cite{Kay2022}
and across phase boundaries~\cite{Bo2021}.
Recently, first-passage problems arose 
in the field of stochastic thermodynamics~\cite{Seifert2012} 
where they give insight into the
statistics of stochastic currents~\cite{currents,Singh2019}, 
thermodynamic entropy production~\cite{decision,stopping,Falasco2020,Neri2022}, 
and dynamical activity~\cite{Garrahan2017,Hiura2021} 
in systems driven far from equilibrium.
Lastly, we remark that first-passage phenomena are
intimately tied to the
statistics of extremes~\cite{Kac,Schehr,Satya_records,Hartich_JPA}, 
and their mathematical description has been extended to capture quantum systems~\cite{Eli_QM1,Eli_QM}, 
additive functionals of stochastic 
paths~\cite{Kearney2005,Kearney2007,Kearney2014,Kearney2021,Majumdar2021,Singh2022}, 
intermittent targets~\cite{MercadoVsquez2019,Kumar2021,Spouge1996,Scher2021,Kumar2023},
active particles~\cite{Woillez2019,Mori2020,DiTrapani2023},
non-Markovian 
dynamics~\cite{Hnggi1983,Hanggi1985,Gurin2016}, and resetting processes~\cite{Evans2011,Kusmierz2014,Reuveni2016,Pal2017,Pal2019,Evans2020,Besga2020,TalFriedman2020,DeBruyne2020,DeBruyne2022,Stojkoski2022},
\blue{including soft resets \cite{Deng2022}}.

Unfortunately, the exact first-passage time density and thus the full statistics
are typically known only for rather elementary examples and simple geometries
(e.g., see Ref.~\cite{redner2001guide}), 
and therefore remains a grand theoretical challenge on its own. The
long-time behavior of the first-passage time density is generally
somewhat better accessible but still requires substantial efforts \cite{Hartich_JPA,Hartich_2018,Hartich_2019}.  
A more accessible, albeit still demanding, 
statistic 
is the \emph{mean first-passage time}. 
Remarkably, this simplest statistic is often a highly non-trivial
quantity, especially in cases where the dynamics span a broad range
of timescales.
Here, some trajectories almost directly reach the target and
contribute to the short-time behavior, whereas others survive much
longer and feed into the long-time behavior \cite{Godec_PRX}.
This makes an estimation 
of the mean first-passage time or its inverse (i.e., the kinetic rate constant)
from experimental or simulation data a complex problem.
The central theme of this chapter is thus 
how to efficiently control the uncertainty of such empirically 
derived estimates of the mean first-passage time, especially in the context
of small sample-sizes.

\begin{figure}[tbp]
\centering
\includegraphics[scale=1]{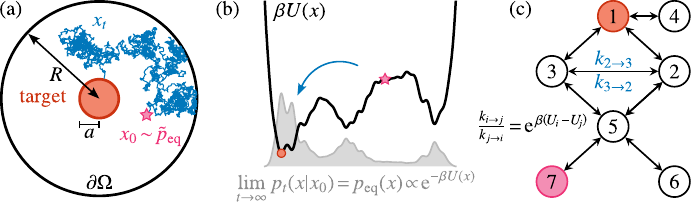}
\caption{Schematic of target search processes for ergodic reversible Markov dynamics.
Diffusive dynamics in 
(a) $d$ dimensional spherical domains (here $d=2$)
with reflecting boundary $\partial\Omega$ and 
(b) arbitrary one-dimensional confining potential landscapes $U(x)$.
(c) Markov jump dynamics on a discrete network state-space
with transition rates that obey detailed balance.
Search processes are initialized from $x_{t=0}$ (magenta), which is drawn from
the stationary density $\tilde{p}_{\rm eq}(x)$ 
and we consider the first-passage time $\tau$ to reach the target (red).
}
\label{Fig1}
\end{figure}

We will focus on the first-passage properties of general target searches described
by reversible Markovian dynamics  
in discrete or continuous state-spaces (see Fig.~\ref{Fig1}). 
In particular, we will discuss Markov jump processes with arbitrary
transition-rate matrices and effectively one-dimensional diffusion
processes in arbitrary confining potentials $U(x)$ that also include
radial representations of hyper-spherically symmetric
diffusion processes in $d$ dimensions
\blue{where the spurious drift in radial direction
allows the mapping to a 
one-dimensional geometric free energy 
of purely entropic origin (e.g., see \cite{Godec_PRX}).}

The chapter is structured as follows.
We begin in Sec.~\ref{sec:fpt_generic_properties} 
with an overview of general first-passage concepts. In Sec.~\ref{sec:fpt_spectral}
we outline the basics of spectral theory of first-passage processes.
Two illustrative examples are worked out in 
Secs.~\ref{sec:markov_network} and~\ref{sec:diffusion_example}.
Section~\ref{sec:mfpt_inference} introduces the challenges associated with 
estimating mean first-passage times from observations, and
Sec.~\ref{sec:deviation_bounds} explores \emph{concentration
inequalities} for bounding deviations 
of inferred estimates from the true mean first-passage time,
applicable for all sample sizes. 
Building on these results, 
Section~\ref{sec:confidence_intervals} demonstrates how they can be
applied to construct 
performance guarantees that do \emph{not} hinge on asymptotic assumptions.
In Section~\ref{sec:beyond_mfpt} we move beyond mean first-passage times
and showcase results that bound deviations 
of minimum and maximum first-passage times in samples of arbitrary sizes.
These findings are relevant in situations when 
the mean is \emph{a priori} an insufficient statistic and problems
involving multiple searchers. 
The chapter concludes in Sec.~\ref{sec:conclusions} with a discussion
and an outlook.

\section{First-passage fundamentals: a spectral-theoretic perspective}
\label{sec:fpt_foundations}
In this section we present some 
background, by introducing elementary
concepts of first-passage theory of target search.
Of particular importance is hereby the mathematical structure 
encoded in a spectral decomposition of first-passage time densities,
which will be the essential building block of derivations throughout the chapter.

\subsection{Generic concepts in first-passage theory}
\label{sec:fpt_generic_properties}
Let $x_t$ denote
the abstract trajectory of a ``searcher'' in a state space $\Omega$; for instance, it may correspond 
to the position coordinate of a diffusive particle confined within 
a spherical domain (see Fig.~\ref{Fig1}a),
a particle exploring an arbitrary confining potential landscape $U(x)$ (see Fig.~\ref{Fig1}b),
or a network-based representation of 
protein configurations in the search of the native 
conformational state during folding (see Fig.~\ref{Fig1}c).
We suppose that $x_t$ follows a reversible Markovian equation of motion. 
A more precise formulation of reversible
dynamics will be given at the end of this section. The description
will be kept ``light'' and informal.

The {time} required for the searcher to make its \emph{first} encounter with some predefined
target $a\in\Omega$ starting from an initial condition distributed
according to the
density $p_0$ then corresponds to the first-passage time $\tau$, formally defined as, 
\begin{align}
\tau = \inf_t \left[t |x_t = a, p_{0}(x_0) \right].
\end{align}
The stochastic nature of the underlying dynamics inherently
renders the first-passage time a random variable that not only depends on the location of the target $a$ 
but also the initial position $x_{t=0}$ (drawn from $p_0(x)$).
The complete statistics of $\tau$ is encoded 
in the so-called survival probability
\begin{align}
S_a(t|x_0) \equiv \mathbb{P}(\tau\geq t),
\label{eq:SurvProb}
\end{align}
the probability that the searcher has 
\emph{not} found the target by time $t$, i.e., the first-passage
event has not yet occurred by $t$. The normalized probability flux
into the target $a$ corresponds to
the first-passage time density 
$\wp_a(t|x_0)=\langle \delta(t-\tau[\{x_{t'}\}]\rangle$,
where $\langle\cdot\rangle$ 
represents the average over all first-passage paths $\{x_{t'}\}_{0\leq t'\leq\tau}$, i.e., 
paths that hit $a$ only once.
By the fundamental theorem of calculus, the survival probability $S_a(t|x_0)$ in Eq.~\eqref{eq:SurvProb} and
the first-passage time density $\wp_a(t|x_0)$ are related via
\begin{align}
\wp_a(t|x_0) = - \frac{\partial}{\partial t} S_a(t|x_0).
\label{eq:surv_derivative}
\end{align}

\newpage
\begin{backgroundinformation}{Assumptions about the dynamics}
We assume $x_t$ to be a strongly ergodic
time-homogeneous Markov process on a continuous or discrete state
space $\Omega$ with invariant (equilibrium) density $p_{\rm eq}(x)$.
The transition probability density $p_t(x|x_0)$ to find
$x_t$ at $x$ at time $t$, given that it started from $x_0$, is given by
$p_t(x|x_0)=\e^{\hat{L} t} \delta_{x_0}(x)$
where $\delta_{x_0}(x)$ denotes either the Dirac or Kronecker delta, 
depending on the specific state-space being considered, and the generator $\hat{L}$ is a linear reversible (dissipative)
operator that is essentially self-adjoint (i.e., it is self-adjoint in
some basis; e.g., the operator $\sqrt{p_{\rm eq}(x)}^{-1}\hat{L}\sqrt{p_{\rm eq}(x)}$ is self-adjoint).

In our context, $\hat{L}$ can take the form of
a Markov-rate matrix or, equivalently, an effectively 
one-dimensional Fokker-Planck operator
with discrete spectrum and real non-positive eigenvalues; obeying detailed balance in both cases. 
\blue{The effectively one-dimensional character of the process
is crucial, as the theory in Sec.~\ref{sec:fpt_spectral} 
builds on the renewal theorem \cite{Siegert1951,Hartich_2018,Hartich_2019}.
For the latter to hold, a process starting from some $x_0$ 
\emph{must}, before reaching the final state $x$, pass through the point-like
target set $a$ that, however, has zero measure in $d\geq 2$. If we
consider absorbing sets with positive measure, the renewal structure
breaks down unless the symmetry allows a reduction to an effectively
one-dimensional Markov dynamics. The results may
even hold without this requirement, but are not covered by our proof.}
As we assume $x_t$ to be ergodic, $p_t$ reaches Boltzmann-Gibbs equilibrium 
for long times  $\lim_{t\to\infty}p_t(x|x_0) = p_{\rm eq} \propto \exp(-\beta U)$
with thermal energy $\beta^{-1}=k_{\rm B}T$.
Lastly, a discrete and real spectrum of eigenvalue is ensured
by assuming that $\hat{L}$ is 
(i) either bounded (Fig.~\ref{Fig1}c)
(ii) $\Omega$ is finite with reflecting boundary $\partial\Omega$ 
(Fig.~\ref{Fig1}a)
or (iii) that $\beta U(x)$ is sufficiently confining 
(Fig.~\ref{Fig1}b).

The processes discussed above are \emph{relaxation} processes that 
conserve probability. In the first-passage setting one has to additionally
introduce the absorbing target $x=a$, which formally modifies the
generator $\hat{L}\to\hat{L}_a$.
For discrete-state dynamics the generator is obtained
by removing all transitions out of the absorbing target, i.e., 
$\hat{L}_a = \hat{L} - \hat{L}|a\rangle\langle a|$.
Here $|a\rangle$ is a vector containing zero entries except
an entry of unity 
at the $a$-th position.  
In a continuous state-space $L_a$ remains the Fokker-Planck operator, however, 
the subscript $a$ now denotes an absorbing (Dirichlet) boundary condition $p_t(a,t|x_0)=0$.
\end{backgroundinformation}

\subsection{Spectral decomposition of first-passage processes}
\label{sec:fpt_spectral}
The reversibility of the underlying Markov process and the
finite/confining domain  ensure 
that the first-passage generator $\hat{L}_a$ has
a spectral expansion in a bi-orthogonal eigenbasis
\begin{equation}
\hat{L}_a = - \sum_{k\ge 1} \mu_k |\phi_k^{\rm R}\rangle\langle\phi_k^{\rm L}|
\end{equation}
in terms of the eigenvalues $\mu_k>0$ and right $|\phi_k^{\rm R}\rangle$
and left $\langle\phi_k^{\rm L}|$ real eigenfunctions, respectively.
Without loss of generality we assume the ordering $\mu_k \leq  \mu_{k+1}$, where $\mu_1 >0$.
For more details on the spectral-theoretic description of relaxation
and first passage see~\cite{Hartich_2018,Hartich_2019}.

The first-passage time density $\wp_a(t|x_0)$
to reach the absorbing target at $x=a$ starting from
$x_0$ in turn also admits a spectral representation
of the form~\cite{Hartich_2018,Hartich_2019}
\begin{equation}
\wp_a(t|x_0) = \sum_{k \geq 1} w_k(x_0) \mu_k \e^{-\mu_k t}.
\end{equation}
Each eigenvalue $\mu_k$ corresponds to a distinct first-passage time-scale via $1/\mu_k$,
and the contribution of each time-scale is governed by the
corresponding (generally not necessarily positive) first-passage
weight $w_k(x_0)$. The weights are normalized according to $\sum_k
w_k(x_0)=1$. Moreover, first-passage weights explicitly depend on the
initial condition $x_0$
and for pseudo-equilibrium initial conditions $x_0 \sim \tilde{p}_{\rm
  eq}(x)$  \blue{(i.e., equilibrium but excluding---effectively disregarding---the target; see below)}
one always finds $w_k^{\rm eq}\geq 0, \forall k$ \cite{bebon2023controlling}.
Furthermore, it is important to highlight that both, $\mu_k(a)$ and $w_k(x_0,a)$, depend on the location of
the absorbing target $a$, which we strictly account for, but omit in
the notation for the sake of readability.
The
survival probability Eq.~\eqref{eq:surv_derivative}
can equally be expressed in a spectral form,
\begin{equation}
S_a(t|x_0) = \sum_{k\geq 1} w_k(x_0) \e^{-\mu_k t},
\end{equation}
and the $m$-th first-passage moment is computed via
$\langle \tau^m \rangle = m! \sum_{k\geq 1}w_k(x_0) / \mu_k^m$.

\begin{warning}{Pseudo-equilibrium initial conditions}
We stress that we will in all subsequent discussions 
assume that $x_{t=0}$ is drawn from the pseudo-equilibrium density 
$p_0(x) = \tilde{p}_{\rm eq}(x)$.  This has important consequences for
$w_k(x_0)$ (and hence $\wp_a(t|x_0)$ and
$S_a(t|x_0)$). These  pseudo-equilibrium initial conditions apply in a large number of 
settings~\cite{Olivier_NC, Bnichou2014, Mattos2012, chevalier2010first, meyer2011universality, bowman2013introduction,noe2019boltzmann,Braun2019}
since, for example, 
$x_0$ usually cannot be controlled in practice
and is thus effectively sampled from the stationary density.
For a discussion of general initial conditions see~\cite{bebon2023controlling}.

In contrast to the equilibrium density $p_{\rm eq}(x)$
for standard relaxation processes (see Fig.~\ref{Fig1}b in gray), 
\blue{the absorption process will reach the target with probability 1 as $t\to\infty$.}
A corresponding definition of a non-trivial stationary \emph{pseudo-equilibrium} density $\tilde{p}_{\rm eq}(x)$
as an initial condition for the first-passage process (i.e.,
absorption) therefore becomes more nuanced by the presence of the absorbing target $a$.
\blue{
For discrete state-spaces $\Omega_{\rm ds}$ with states $x\in\Omega_{\rm ds}$
the pseudo-equilibrium density is obtained
via a renormalization of $p_{\rm eq}(x)$ by simply excluding the target 
in the sense of 
$\tilde{p}_{\rm eq}(x)\equiv p_{\rm eq}(x)/\sum_{k\neq a}p_{\rm eq}(k)$
for $x\neq a$ and $ \tilde{p}_{\rm eq}(x) =0$ when $x=a$.
The resulting pseudo-equilibrium density therefore 
adopts the target as perfectly reflecting,
effectively disregarding its presence.  
This ensures $S_a(t=0|x_0\sim\tilde{p}_{\rm eq})=1$.
Note that by contrast, in this case the \emph{true invariant
distribution of the absorption  
process} is trivially $p_{\rm inv}^{\rm abs}(x)=\delta_{x,a}$.
The situation is straightforward for continuous state-spaces,
where we have $\tilde{p}_{\rm eq}(x)= p_{\rm eq}(x)$,
since the target $a$ has zero measure.
}
\end{warning}

\subsection{Example 1: Markov jump network}
\label{sec:markov_network}
For illustrative purposes we first consider a Markov process on a
discrete state space 
(see Fig.~\ref{Fig1}c)
as a model of a conformational search of a protein 
for the native folded state $a=1$ (red), 
with the initial condition $x_0$ drawn from the pseudo-equilibrium $\tilde{p}_{\rm eq}(x)$.
The dynamics are encoded in the transition rates $k_{i\to j}$ from state $i$ to $j$
and enter the transition-rate matrix $\hat{L}$ via $L_{ji} = k_{i\to
  j}$ and $L_{ii} = -\sum_{j\neq i}L_{ji}$. 
Since we consider reversible dynamics, $\hat{L}$ obeys detailed balance
$p_{\text{eq},j}/p_{\text{eq},i} = L_{ji}/L_{ij} = \exp[\beta(U_i-U_j)]$,
and the transition rates are related to the (free) energy of states $U_i$.
Computing the weights is readily achieved by diagonalization of
$\hat{L}_a$ and using
$w_k(x_0) = - \langle a|\phi_k^{\rm R}\rangle\langle\phi_k^{\rm L}|x_0\rangle$~\cite{Hartich_2018, Hartich_2019}.
Recall that $w_k^{\rm eq}\geq 0, \forall k$ as a result of the
pseudo-equilibrium initial condition \cite{bebon2023controlling}.

\subsection{Example 2: diffusive target search in a confined spherical domain}
\label{sec:diffusion_example}
As a second example we analyse 
a confined diffusive search of a Brownian particle in a $d=3$ dimensional 
unit sphere with reflecting boundary $\partial \Omega$ at $R=1$
(see Fig.~\ref{Fig1}a).
A perfectly absorbing spherical target with radius $0<a<1$ 
is placed in the center (we take $a=0.1$).
The time-evolution of the 
distance to the absorbing sphere $x_t$
is a confined Bessel process obeying
the \blue{
stochastic Skorokhod equation \cite{Skorokhod1961,McKean1965,Tanaka1979,Lions1984,Grebenkov2019PRE,Grebenkov2020PRL}
\begin{equation}
d x_t = \frac{2}{x_t} dt + \sqrt{2} dW_t + \vmathbb{1}_{\partial \Omega}(x_t) d L_t,
\end{equation}
where $dW_t$ denotes the Wiener increment, i.e., Gaussian white noise
with $\langle dW_t\rangle =0$,
$\langle dW_t dW_{t'}\rangle =\delta(t-t')dtdt'$,
and we set the diffusion constant to unity $D=1$ without loss of generality.
The second term explicitly accounts for (normal) reflections on the
boundary $\partial \Omega$, i.e.,\ that $x_t$ does not leave the domain $\Omega$.
Here, the boundary local time $L_t$ is
a non-decreasing process with $L_0 = 0$, that by construction only
increases when $x_t\in\partial\Omega$, i.e., 
the indicator function $\vmathbb{1}_{\partial\Omega}(x_t)=1$ if $x_t\in\partial\Omega$
and $0$ otherwise.}
Note that the general case of a sphere with radius $R$ with any $0<D<\infty$
is easily recovered by expressing time in units $R^2/D$.
It is not difficult to show that the pseudo-equilibrium weights are given by
\begin{equation}
w_k^{\rm eq}=\frac{2}{\mu_k}\frac{3a^2}{1-a^3}\frac{\tan[(1-a)\sqrt{\mu_k}]+\frac{1}{\sqrt{\mu_k}}}{(1-a)\tan[(1-a)\sqrt{\mu_k}]-\frac{a}{\sqrt{\mu_k}}},
\end{equation}
and the first-passage rates $\mu_k$ are solutions of the transcendental equation $\sqrt{\mu_k}=\tan([1-a]\sqrt{\mu_k})$.

\section{Inference of the mean first-passage time}
\label{sec:mfpt_inference}
From a practical point of view,
the inference of the mean first-passage time $\mtau$ from experimental
or simulation data 
via the direct 
\emph{unbiased}\footnote{The bias of an estimator (e.g., the sample-mean)
is defined as the difference between its expectation
and the true value. Correspondingly, estimators with zero bias
are said to be unbiased \cite{hogg1995introduction}.}
sample-mean estimator 
that we call \emph{empirical first-passage time} $\ntau\equiv n^{-1}\sum_{i} \tau_i$,
generally poses a serious challenge, especially in situations where only a
limited number $n$ of first-passage events $\tau_i$ are available. The
sub-sampling may be
due to, e.g., practical constraints in experimental setups
or limited computational resources. 
Hence, sample numbers in the
range from 1-10~\cite{lindorff2011fast,Adelman2013,Bert,Mostofian2019,Mehra2019,Militaru2021} 
or sometimes up to 100~\cite{Rondin2017} are quite common.
Clearly, when dealing with sub-sampling issues,
the inferred estimate is afflicted by large uncertainties,
especially when first-passage times are distributed 
across many time-scales and
extreme values substantially contribute to (and potentially skew) the sample mean.
Understanding the fluctuations of $\ntau$ for any $n$
is thus crucial but 
challenging, since limited sample-sizes lead to non-Gaussian
errors in turn
rendering uncertainty quantification not amendable to standard error-analysis techniques.

As detailed in the following section,
we recently addressed the statistical deviations of the inferred empirical $\ntau$ from the actual mean first-passage 
time $\mtau$ by bounding the probability of such deviations
in the small-sample regime~\cite{bebon2023controlling}.
Our approach rests on
a non-asymptotic description of the \emph{concentration-of-measure
phenomenon}. 

\subsection{Bounding the probability of deviations}
\label{sec:deviation_bounds}
The study of the concentration behavior of the sample-mean (here $\ntau$) around
the true mean (here $\mtau$)
has a long tradition in probability theory.
For example, in the asymptotic limit $n\to\infty$
the (weak or strong) law of large numbers establishes that 
$\ntau$ converges to $\mtau$ (in probability or almost surely)
and the central limit theorem asserts that
$\sqrt{n}(\ntau-\mtau)$ 
converges to a normal distribution with mean 0 and variance
$\sigma^2$. 
Similar statements for finite $n$ 
pose a much greater challenge that requires a more intricate analysis.
\blue{On a related note, we remark that understanding finite-sample
  effects is important beyond statistical inference, as these manifest in diverse 
contexts such as, e.g., 
molecule-number fluctuations in chemical reactions \cite{Metzler2015}.}

\begin{figure}[tp]
\centering
\includegraphics[scale=1]{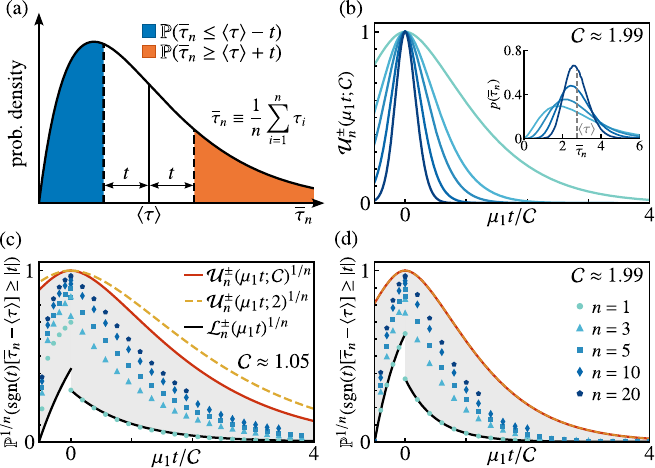}
\caption{Non-asymptotic concentration 
\blue{and sample-size effects} of empirical first-passage times $\ntau$
  around the true mean $\mtau$.
(a) Schematic probability density of $\ntau$ inferred from a (small)
  sample of $n$ realizations. Fluctuations are quantified by tail
  probabilities of deviations of $\ntau$ from $\mtau$ by more than $t$
towards the right $\mathbb{P}(\ntau\geq\mtau +t)$ 
or the left $\mathbb{P}(\ntau\leq\mtau -t)$ and are
shown in orange and blue, respectively.
\blue{(b) Dependence of the upper bounds $\mathcal{U}_n^\pm(\mu_1 t; \C)$ on sample size $n$
for the confined Brownian search in $d=3$. Bounds quantify how the probability of 
deviations from the sample mean drastically diminishes as $n$ increases (bright to dark; see (d)).
Inset: corresponding empirical histogram of the sample mean $\ntau$ for different $n$ values.}
(c,\,d) Deviation probabilities and corresponding bounds
for (c) a Markov network representation of protein folding \blue{($\C\approx 1.05$)} 
and (d) a spatially confined Brownian search \blue{($\C\approx 1.99$)} 
as introduced in Fig.~\ref{Fig1}.
Probabilities are scaled as $\mathbb{P}^{1/n}({\rm sgn}(t)[\ntau
  -\mtau]\geq |t|)$; right tail areas are shown for $t>0$ and left for $t<0$, respectively.
Lower $\mathcal{L}_n^\pm(\mu_1 t)^{1/n}$ and upper bounds $\mathcal{U}_n^\pm(\mu_1 t; \C)^{1/n}$ are
depicted as black and red lines, respectively, 
and the model-free bounds $\mathcal{U}_n^\pm(\mu_1 t;2)^{1/n}$
are shown as yellow dashed line.
Corresponding deviation probabilities obtained by numerical
simulations as a function of $t$ for different fixed $n$ 
are denoted by symbols.
\blue{Yellow and red curves coincide in (d) since $\C\approx 1.99$.}}
\label{Fig2}
\end{figure}

Returning to statistics, to assess the fluctuations of the inferred $\ntau$ around the true mean $\mtau$ 
in more general situations (i.e., small $n$) where 
the sample-average $\ntau$ is \emph{not yet}
tightly concentrated around $\mtau$ 
and its probability density\footnote{Recall that the sample-mean $\ntau$ is a random variable on its own as every set of $n$ i.i.d.~recorded first-passage events $\{\tau_1, \tau_2, \ldots, \tau_n \}$ gives rise to a different estimate $\ntau$.} 
is not close to a Gaussian (see Fig.~\ref{Fig2}a),
we examine
deviation probabilities $\mathbb{P}(\ntau \geq \mtau +t)$ and $\mathbb{P}(\ntau\leq\mtau -t)$, 
respectively.
These reflect the probability that the empirical first-passage time $\ntau$, 
estimated from a sample of
$n\geq 1$ recorded first-passage events, deviates from 
the true mean $\mtau$ by more than some value $t$ in either direction,
as illustrated in Fig.~\ref{Fig2}a by the colored regions. These probabilities gauge
the likelihood and magnitude of fluctuations in the estimate $\ntau$ 
regardless of sample-size; however they cannot be obtained in a
generic fashion since the probability density of $\ntau$ is 
not known.

Here, we describe a solution of this problem in terms of the recently 
derived general upper bounds, so-called concentration inequalities,
$\mathbb{P}(\pm[\ntau-\mtau]\geq t) \leq \mathcal{U}^\pm_n(t)$, which
do \emph{not} require the knowledge of the distribution of $\ntau$.
Notably, these bounds are valid for all values $n\geq 1$ and $t\geq 0$ and
their most conservative version is
independent of any specific details about the underlying dynamics.
From a technical perspective they are derived from a bounding
technique referred to as the 
\emph{Cram\'er-Chernoff method}~\cite{boucheron2013concentration}
which is summarized briefly at the end of this section.

A complete derivation with all mathematical details is beyond the scope of this
chapter (see \cite{bebon2023controlling}), 
we therefore only outline the key concepts.
The procedure starts from Chernoff's inequality $\mathbb{P}(\pm [\tau-\mtau]\geq
t)\leq\e^{-\lambda t +\psi_{\pm (\tau-\mtau)}(\lambda)}$, where $\psi_X(\lambda)$
denotes the cumulant generating function of a random variable $X$ (see below), and essentially involves two main steps.
First, we prove an upper bound on the cumulant generating function
$\ln\langle\exp[\pm\lambda (\tau -\mtau)]\rangle \leq \phi_\pm(\lambda)$, $\lambda\in\mathbb{R}^+$, 
in the form of
\begin{align}
\phi_{+}(\lambda)
=\frac{\lambda^2}{2}\frac{\mu_1^2\langle\tau^2\rangle}{1-\lambda/\mu_1}
\quad\text{and}\quad
\phi_{-}(\lambda)
=\frac{\lambda^2}{2}\frac{\mu_1^2\langle\tau^2\rangle}{1-(\lambda/\mu_1)^2},
\label{eq:phi_bound}
\end{align}
for the \blue{right tail ($\tau\geq\mtau$; index $+$) and left tail ($\tau <\mtau$;  index $-$), respectively.}
In the second step we carry out a
Legendre transform 
$\phi_\pm^\ast(t)\equiv \sup_\lambda[\lambda t - \phi_\pm(\lambda)]$ to
optimize the bound. Considering the statistical independence of the $\tau_i$, 
the concentration bounds are found to have an exponential form 
$\mathbb{P}(\pm[\ntau-\mtau]\geq t)\leq \exp(-n\phi^\ast_\pm(t))$.
Carrying out these steps  we obtain the announced upper bounds~\cite{bebon2023controlling} 
\begin{align}
&\mathbb{P}(\ntau-\mtau\geq t)\leq\exp(-n\C h_+(\mu_1 t/\C))\equiv\mathcal{U}_n^+(\mu_1 t;\C)  \quad && 0\leq t\leq\infty
\nonumber
\\
&\mathbb{P}(\mtau-\ntau\geq t)\leq\exp(-n\C h_-(\mu_1 t/\C))\equiv\mathcal{U}_n^-(\mu_1 t;\C) \quad && 0\leq t\leq\mtau,
\label{eq:concentration_inequality}
\end{align}
where we introduced $\C\equiv\mu_1^2\langle\tau^2\rangle$ 
and defined
the auxiliary functions
\begin{align}
h_+(u)\equiv 1+u-\sqrt{1+2u}
\quad\text{and}\quad
h_-(u)\equiv \Lambda(u)u- \frac{1}{2}\frac{\Lambda(u)^2}{1-\Lambda(u)^2},
\end{align}
with 
\begin{align}
&\Lambda(u)\equiv\frac{1}{2}\left[g(u)-\sqrt{4+2/g(u)-g(u)^2}\right]
\\
&g(u)\equiv\frac{2}{\sqrt{3}}\left\{1+2\cosh\left[\frac{1}{3}\arccos\left(1+\frac{3^3}{2^7u^2}\right)\right]\right\}^{1/2}.
\end{align}
It is worth noting that, for the bounds presented in
Eq.~\eqref{eq:concentration_inequality}, the underlying details of the dynamics
manifest solely via the system-dependent constant $\C$.
Remarkably, however, it can be shown that 
for pseudo-equilibrium initial 
conditions ($x_0\sim\tilde{p}_{\rm eq}$; see end of
Sec.~\ref{sec:fpt_spectral}) this constant falls within the range $0\leq\C\leq 2$~\cite{bebon2023controlling}.
This insight can be leveraged to directly establish 
\begin{equation}
\mathcal{U}_n^\pm(\mu_1 t;\C)\leq\mathcal{U}_n^\pm(\mu_1 t;2)\equiv\mathcal{U}_n^\pm(\mu_1 t).
\end{equation}
While these bounds tend to be more conservative 
(i.e., when the actual value of $\C$ significantly deviates from 2
one may still take $\C=2$ but the bound becomes less sharp),
they are entirely model-independent.
In the case of general initial conditions $p_0(x)\neq \tilde{p}_{\rm
  eq}$, $\C$ becomes replaced with a different constant (for details see \cite{bebon2023controlling}).

At this point we emphasize that
the magnitude of deviations, in other words, the ``error'', 
is naturally parameterized by the dimensionless variable $\mu_1 t$
as evident from the argument of $h_\pm(\mu_1 t /\C)$ in
Eq.~\eqref{eq:concentration_inequality}. 
Therefore, considering errors in units $1/\mu_1$, $\pm\mu_1(\ntau-\mtau)$, 
there is no need to specify or know $\mu_1$ (i.e., the longest time-scale) itself.
This relative error normalized by the ``observable largest value'' of first
passage events $1/\mu_1$  thus quantifies the deviation of  the
inferred empirical mean first-passage time $\ntau$
from the true mean $\mtau$ relative to the characteristic 
(maximal, ``extreme-case-scenario''\footnote{The time-scale $1/\mu_1$
sets a ``cut-off''  on the range of observable
first-passage events; events longer than $\mathcal{O}(1)\times 1/\mu_1$ are exponentially unlikely.}) 
time-scale of the system.
In this context deviation bounds~\eqref{eq:concentration_inequality}
state that, for any sample size $n$, the probability \blue{$\mathbb{P}(\mu_1 [ \ntau-\mtau] \geq \mu_1 t)$}
of observing a relative error larger than a specified (dimensionless) value $\mu_1 t$ 
(e.g., say $\mu_1 t = 0.1$ or 10\%)
is lower that the corresponding upper bound $\mathcal{U}_n^\pm(\mu_1 t)$.

The validity of concentration 
bounds~\eqref{eq:concentration_inequality} \blue{and sample-size effects are}
illustrated in Fig.~\ref{Fig2}b-d for the 
target-search examples introduced earlier.
To make the comparison, 
we evaluated ``exact'' empirical deviation probabilities (symbols) obtained from 
extensive sampling of $\ntau$ ($10^{11}$ repetitions) for various fixed values of $n$ 
\blue{(see inset Fig.\ref{Fig2}b for histograms of $\ntau$)}.
For visualization purposes,
we formally let $t\to -t$ for the left tails, allowing us to depict both, the right and left deviation bounds, in a single plot 
over the support $[-\mtau, \infty)$.
\blue{Illustrated in Fig.~\ref{Fig2}b, deviation bounds quantify an exponential concentration rate of $\ntau$
around $\mtau$ as $n$ increases.}
In panels (c-d) we additionally employed the scaling $\mathbb{P}^{1/n}$ 
to collapse the results onto a single master curve for all values of $n$.
Data points approach the upper bound as $n$ increases and
as anticipated, the model-independent bound $\mathcal{U}_n^\pm(\mu_1 t;2)$ (yellow) holds
universally but tends to be more conservative.

Finally, we mention that in \cite{bebon2023controlling} we also
derived corresponding lower bounds $\mathcal{L}^\pm_n$, for the sake of completeness shown in black,
\blue{
\begin{align}
&\mathbb{P}(\ntau -\mtau \geq t)\geq \left(w_1 \e^{-\mu_1(\mtau +t)}  \right)^n \equiv \mathcal{L}_n^+(\mu_1 t)
\nonumber
\\
&\mathbb{P}(\mtau -\ntau \geq t)\geq \left(1-\e^{-\mu_1(\mtau -t)}  \right)^n \equiv \mathcal{L}_n^-(\mu_1 t),
\label{eq:lower_bound}
\end{align} 
by again leveraging the spectral analysis discussed in Sec.~\ref{sec:fpt_spectral}
in combination with ideas from extreme-value theory.
Specifically, the searcher is assumed to start 
at $x_0$ drawn from pseudo-equilibrium  $\tilde{p}_{\rm eq}(x)$.
A detailed discussion of lower bounds \eqref{eq:lower_bound}---explicitly demonstrating 
the existence of an intrinsic noise-floor---is beyond the scope of this chapter, 
as these do not play a role in the
construction of performance guarantees presented in the following section.
Nevertheless, these non-trivial lower bounds underscore the critical importance
of a robust assessment of uncertainty of the observable $\ntau$, which we explore next.
}

\begin{figure}[tbp]
\centering
\includegraphics[scale=1]{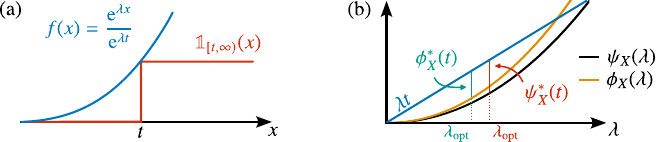}
\caption{Sketch of the Cram\'er-Chernoff method.
(a) Schematic of the inequality in Eq.~\eqref{eq:inequality_1}.
(b) Illustration of the Legendre transform
of $\psi_X(\lambda)$ (black) and $\phi_X(\lambda)$ (yellow).
The optimal $\lambda_{\rm opt}$ maximizes
the difference with $\lambda t$ (blue);
$\psi_X^\ast(t)$ (red) and $\phi_X^\ast(t)$ (green)
denoting the corresponding Legendre transform (Eq.~\eqref{eq:legendre}), respectively.
Note that $\phi_X(\lambda)\geq \psi_X(\lambda)$ implies 
$\phi_X^\ast(t)\leq\psi_X^\ast(t)$.}
\label{Fig3}
\end{figure}
\begin{backgroundinformation}{Cram\'er-Chernoff method in a nutshell}
Here we present a short overview of the Cram\'er-Chernoff method,
for more details see, e.g., \cite{boucheron2013concentration}.
Consider a real-valued positive random variable $X$, in our context either $\tau - \mtau$ for the right tail or $\mtau - \tau$ for the left tail.
For any $\lambda \in \mathbb{R}^+$, we begin with the obvious inequality (see Fig.~\ref{Fig3}a):
\begin{equation}
\vmathbb{1}_{[t, \infty)} \leq \frac{\e^{\lambda X}}{\e^{\lambda t}}\quad \forall X\geq 0,
\label{eq:inequality_1}
\end{equation}
\blue{where $\vmathbb{1}_{A}$ denotes the indicator function of set $A$.}
Taking expectation over $X$, $\langle \cdot\rangle$, on both sides, we
immediately obtain 
\begin{equation}
\mathbb{P}(X\geq t)\leq \e^{-\lambda t}\langle \e^{\lambda X}\rangle
=\e^{-\lambda t +\psi_X(\lambda)},
\label{eq:cramer_chernoff1}
\end{equation}
where $\psi_X(\lambda)\equiv\ln\langle \e^{\lambda X}\rangle$ denotes the cumulant generating function of $X$.
Since this inequality holds for all values $\lambda\geq 0$, one can select $\lambda$ in a way
that optimizes the upper bound, achieved by computing the 
Cram\'er-Legendre transform (see Fig.~\ref{Fig3}b)
\begin{equation}
\psi^\ast_X(t) = \sup_{\lambda\in\mathbb{R}^+}[\lambda t - \psi_X(\lambda)].
\label{eq:legendre}
\end{equation}
By Eq.~\eqref{eq:cramer_chernoff1} we thus obtain an upper bound on the tail probability
according to
\begin{equation}
\mathbb{P}(X\geq t)\leq \e^{-\psi_X^\ast(t)}.
\end{equation}
The strength of the method lies in its extension to sums
$\overline{X}=X_1 + \ldots + X_n$ of $n$ i.i.d.~real-valued random variables 
$X_1, \ldots X_n$.
Due to independence, we have $\psi_{\overline{X}}(\lambda) =
n\psi_X(\lambda)$, leading to
$\psi^\ast_{\overline{X}}(t) = n\psi^\ast_X(t/n)$,
in turn allowing one to write
\begin{equation}
\mathbb{P}(\overline{X}/n \geq t) \leq \e^{-n\psi^\ast_X(t)},
\end{equation}
which underscores the evident connection to sample-averages like $\ntau$.
We remark that in our case, the Legendre transform was computed 
not directly for $\psi_X(\lambda)$ but rather for $\phi_X(\lambda)\geq\psi_X(\lambda)$ 
(Eq.~\eqref{eq:phi_bound}), 
implying that $\psi_X^\ast(t) \geq \phi_X^\ast(t)$ (see Fig.~\ref{Fig3}b).
\end{backgroundinformation}

\subsection{Uncertainty quantification}
\label{sec:confidence_intervals}
Having established the importance of assessing uncertainty in 
estimates derived from undersampled data, 
we are now confronted with the challenge of constructing reliable
performance guarantees  
for the kinetic inference of empirical first-passage times $\ntau$. 
Unfortunately, for the task at hand many standard procedures fall short as they
are grounded in asymptotic theory that necessitates 
a  large sample size ($n\to\infty$). 
This requirement, clearly, is practically unattainable,
making their applicability 
in sub-sampled regimes, by definition, problematic.
Likewise, complementary approaches to statistical inference via 
Bayesian methods (see e.g., Ref. \cite{gelman1995bayesian}) 
face challenges when dealing with limited data, 
as the influence of prior distributions must be carefully considered.
A sensitivity to the choice of a prior potentially results in
``biased'' prior-dependent uncertainty estimates, even 
in the large-sample limit.

This apparent
lack of accurate finite-sample guarantees
highlights the pressing need for reliable non-asymptotic error estimation 
techniques---e.g., in the form of confidence intervals---with correct 
finite-sample coverage probabilities 
to assess the quality
of inferred estimates.
Luckily, the concentration bounds~\eqref{eq:concentration_inequality} 
hold for any $n$,
thus providing a well-suited candidate for addressing the challenges of the small-sample regime. 
As we will outline next, they provide a  framework to quantify 
the discrepancy between the estimate $\ntau$ and the true parameter $\mtau$
in terms of non-asymptotic 
``with high probability'' guarantees.

\begin{figure}[tbp]
\centering
\includegraphics[scale=1]{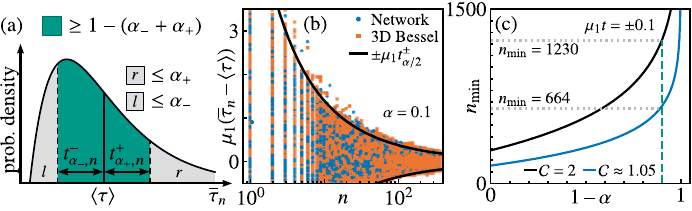}
\caption{Uncertainty quantification of the sample-mean $\ntau$ for any sample-size $n$.
(a) Probability that the sample-mean $\ntau$ lies within an interval
of $[-t_{\alpha_-,n}^-, t_{\alpha_+,n}^+]$ around $\mtau$ with a probability of 
at least $1-(\alpha_- + \alpha_+)$.
(b) Model-free 90\% confidence region (i.e., $\alpha=0.1$) that
the error in units $1/\mu_1$ remains within $\pm 10\%$ as a function
of sample-size $n$.
(c) Minimal-sample size $n_{\rm min}$ required 
to ensure that the relative error $\mu_1(\ntau-\mtau)$
does not exceed $\pm 10\%$ 
as a function of confidence level $1-\alpha$; a confidence of at least $90\%$
is shown as dashed line.
}
\label{Fig4}
\end{figure}

A construction of such performance guarantees is straightforward and
easily implementable.
By setting $\mathcal{U}_n^\pm(\mu_1 t_{\alpha_\pm, n}^\pm)=\alpha_\pm$ for a user-defined
acceptable left and right error probability of $\alpha_\pm$,
we promptly obtain the implicit definition of
a confidence interval $[-t_{\alpha_-,n}^-, t^+_{\alpha_+,n}]$ in 
the form
\begin{equation}
\mathbb{P}(-t_{\alpha_-,n}^-\leq \ntau-\mtau\leq t_{\alpha_+,n}^+)\geq 1 - (\alpha_- + \alpha_+)\equiv 1-\alpha
\label{eq:confidence_interval}
\end{equation}
by performing an ordinary union bound.
In other words, Eq.~\eqref{eq:confidence_interval} asserts
that with a high probability of at least $1-\alpha$ 
the inferred sample-average $\ntau$ falls within the interval
$[\mtau-t_{\alpha_-,n}^-, \mtau + t_{\alpha_+,n}^+]$ (see green region in Fig.~\ref{Fig4}a). 
For illustrative purposes we present in Fig.~\ref{Fig4}b
the most conservative, universal $90\%$ confidence region
(i.e., $\alpha=0.1$ and $\alpha_\pm = \alpha/2$) for the relative error as a function of the given 
sample-size $n$ for the model systems previously introduced 
in Secs.~\ref{sec:markov_network} and~\ref{sec:diffusion_example}.

The non-asymptotic confidence intervals \eqref{eq:confidence_interval}
are further valuable in addressing the equally important question of how large 
the minimal number of realizations $n_{\rm min}$ has
to be if one wants to ensure 
that the error of the estimate $\overline{\tau}_{n_{\rm min}}$
does \emph{not} exceed a specified threshold with a desired level of confidence---an essential consideration 
in the planning of experimental or computational studies.
Correspondingly, this number is implicitly defined via
\begin{equation}
\mathcal{U}^-_{n_{\rm min}}(\mu_1 t^-_{\alpha_-,n};\C) + \mathcal{U}^+_{n_{\rm min}}(\mu_1 t^+_{\alpha_+,n};\C) = \alpha.
\label{eq:min_samplesize}
\end{equation}
Using our simple examples, we explicitly show in Fig.~\ref{Fig4}c the
minimal sample-size $n_{\rm min}$ necessary
to guarantee a (dimensionless) relative error $\mu_1(\ntau-\mtau)$ of at most $\pm 10\%$
as a function of the confidence level $1-\alpha$
for the Markov network model (blue; $\C\approx 1.05$) 
and compare them with the most conservative model-independent version (black line).
Note that the conservative bound holds for 
\emph{any} setup that is described by Markovian dynamics as we introduced in Sec.~\ref{sec:fpt_generic_properties}
and a reliable inference of the mean (i.e., 90\% confidence level; dashed line in Fig.~\ref{Fig4}c)
typically requires hundreds to thousands of samples.

Collectively, Eqs.~\eqref{eq:confidence_interval} 
and~\eqref{eq:min_samplesize} 
provide a framework for evaluating uncertainties that arise
in the kinetic inference of $\ntau$ when dealing with undersampled data.
The required values of $t_{\alpha_\pm, n}^\pm$ and $n_{\rm min}$ can readily 
be determined using standard root-finding techniques (e.g., bisection method).
We remark that in our discussion, we have chosen equal tail probabilities $\alpha_\pm = \alpha/2$, 
resulting in what are commonly referred to as ``central'' confidence intervals. 
However, in general, two-sided confidence regions are not uniquely defined by their confidence level,
and various other choices exist (e.g., symmetric intervals).
Expanding on this, 
there are situations where only one-sided confidence intervals are 
needed \cite{bebon2023controlling}.
In this case we even explicitly obtain that 
\begin{equation}
\ntau - \mtau \leq -\frac{\ln(\alpha)}{\mu_1 n} + \frac{\sqrt{2} \sqrt{-\ln(\alpha)}}{\mu_1\sqrt{n/\C}}
\end{equation}
with probability of $1-\alpha$.
Correspondingly, one obtains the explicit bound
\begin{equation}
n_{\rm min}^\pm \geq - \frac{\ln(\alpha_\pm)}{\C h_\pm(\mu_1 t/\C)}
\end{equation} 
where $n_{\rm min}^\pm$ now denotes the required sample-sizes to ensure that $\pm(\ntau -\mtau)\leq t$ with 
a probability of at least $1-\alpha_\pm$.

\section{Going beyond the mean first-passage time}
\label{sec:beyond_mfpt}
Thus far we reviewed recent advances in quantifying uncertainty of the
sample-mean $\ntau$ inferred from a
(small) finite number $n$ of experimental or simulated trajectories,
and explored how many samples are needed to ensure
that the uncertainty of $\ntau$ falls below
a predefined threshold value.
However, the fact that possibly multiple relevant time-scales are at play 
(e.g., in ``compact'' search processes~\cite{Olivier_NC,Bnichou2014,Godec_PRX,Godec_SciRep})
introduces an additional layer of complexity
to the uncertainty of $\ntau$. In such situations the mean 
on its own may not be a sufficient statistic (i.e., as the
``characteristic'' time-scale) and \emph{a priori} may not faithfully characterize the entire distribution\footnote{Clearly, for an exponentially distributed $\tau$ the mean $\mtau$ gives complete knowledge since $\wp_a(t)=1/\mtau \e^{-t/\mtau}$.}. 
This calls for a deeper understanding 
of the first-passage time density $\wp_a(t)$, including aspects such as its 
behavior in the short and long-time regimes \cite{Grebenkov_2018_2, Olivier_NC,Bnichou2014,Godec_PRX,Godec_SciRep}. 

Interestingly, assuming that a reliable estimate of the mean is available 
(which can be achieved using the outlined procedure 
designed to handle uncertainty in $\ntau$ even where estimation
becomes challenging, including scenarios involving multiple time-scales), 
we can apply the upper bounds $\mathcal{U}_n^\pm(t)$ to gain insight. 
By inspecting the case $n=1$, i.e., $\ntau =\tau$ and 
$\mathbb{P}(\pm[\tau-\mtau]\geq t)\leq \mathcal{U}^\pm(t)$,
the bounds directly provide knowledge about the tail regions of the first-passage 
time density $\wp_a(t)$ (see Fig.~\ref{Fig2}a) since, in this scenario, 
the mean first-passage time $\mtau$ acts as a ``reference point''.
Understanding these tail regions is crucial, as 
trajectories that survive for long times (right tail) 
explore their environment extensively, and thus retain information
about structural properties.  
Conversely, shorter trajectories (left tail) typically reach their target more directly, 
often following the shortest direct path and reveal intricate features
such as the number of energy minima~\cite{Tolya}.
On top of that, in cases where the mean is not \emph{a priori} representative
the spread of extreme (i.e., minimal and maximal) first-passage times 
provides insight about the range of first-passage events (see below).

Another intriguing and related challenge emerges when moving 
beyond the ``single searcher setting'' to the  first-passage
properties of multiple searchers --- the fastest search time \cite{Hartich_2018,Hartich_Book,Lawley_1,Lawley_2,Lawley_3}. 
Here, the statistics may be dominated by the time it takes the fastest (minimal $\tau$) or slowest (maximal $\tau$) 
out of many searchers to reach a target for the first time.
This phenomenon is particularly evident in various biological systems, 
such as the search process of sperm cells seeking an egg cell~\cite{Yang2015}.

Motivated by these challenges, we now summarize recent progress in
understanding first-passage properties  
in scenarios where the mean first-passage time $\mtau$ alone
falls short in capturing the complete range of the dynamics
or when dealing with multiple 
searchers \cite{bebon2023controlling}.

\subsection{Bounding extreme deviations}
The presence of  many different distinct time-scales is already included in the spectral expansion 
of the first-passage time density $\wp_a(t) = \sum_k w_k^{\rm eq}\mu_k \exp(-\mu_k t)$; 
recall that $w_k^{\rm eq}\geq 0, \forall k$ for pseudo-equilibrium
initial conditions.
Evidently, whenever more than a single weight $w_k^{\rm eq}$ contributes, 
multiple time-scales become relevant.
In such cases, relying solely on the knowledge of $\mtau$ may not suffice for an accurate characterization of the entire dynamics, as $\wp_a(t)$ significantly deviates from a single exponential.
For the following discussion we will make use of the fact that the
presence of multiple time-scales is reflected in the dimensionless factor $\mu_1
\mtau \in(0,1]$, i.e.,
it substantially differs from 1, and the dimensionless parameter
$\C=\mu_1^2\langle\tau^2\rangle$ becomes substantially smaller than 2.

\begin{figure}[tbp]
\centering
\includegraphics[scale=1]{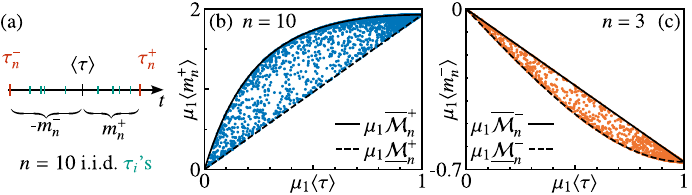}
\caption{Extreme deviations from the mean first-passage time. 
(a) Maximal $\tau_n^+\equiv\max_{i\in[1,n]}\tau_i$ 
and minimal $\tau_n^-\equiv\min_{i\in[1,n]}\tau_i$ first-passage time
in a sample of $n$ i.i.d.~realizations.
(b,\,c) Average maximal ($+$) and minimal ($-$) deviation from 
the mean $\langle m_n^\pm\rangle\equiv\langle \tau_n^\pm -\mtau \rangle$
from extensive computer simulations (symbols) with 
lower $\underline{\mathcal{M}}_n^\pm$ 
(dashed black lines) and upper $\overline{\mathcal{M}}_n^\pm$ bound (black lines)
shown for
$n=10$ and $n=3$, respectively.
Quantities are expressed in units of $1/\mu_1$ and shown as a function of $\mu_1\mtau$.}
\label{Fig5}
\end{figure}

To address the range of these sample-to-sample fluctuations 
of the first-passage time $\tau$
we consider \emph{extreme deviations from the mean}, i.e., 
we are interested how much the maximal $\tau^+_n\equiv \max_{i\in[1,n]} \tau_i$ and
minimal $\tau_n^-\equiv \min_{i\in[1,n]} \tau_i$ 
in a sample of $n$ i.i.d.~observed first-passage times $\tau_i$
deviate from the mean (see Fig.~\ref{Fig5}a) 
on average.
This expected range of extreme deviations 
$\langle m_n^\pm\rangle\equiv\langle\tau_n^\pm -\mtau\rangle$
is the appropriate observable controlling uncertainty when
multiple time-scales are concerned, especially in the small-sample setting, 
as it provides information regarding the spread (i.e., uncertainty) 
of observations.
To this end, we recently derived sharp two-sided ``squeeze'' bounds \cite{bebon2023controlling}
\begin{equation}
\underline{\mathcal{M}}_n^\pm \leq \langle m_n^\pm \rangle \leq \overline{\mathcal{M}}_n^\pm, 
\quad n\geq 1
\end{equation} 
in the form of
\begin{align}
&\overline{\mathcal{M}}_n^+\equiv \sum_{k=1}^n \binom{n}{k}(-1)^{k+1} \frac{(\mu_1\mtau)^k}{\mu_1 k}-\mtau,
&&\overline{\mathcal{M}}_n^-\equiv \mtau\left[\frac{1}{n}-1\right]
\nonumber
\\
&\underline{\mathcal{M}}_n^+ \equiv \mtau \sum_{k=2}^n\frac{1}{k},
&&\underline{\mathcal{M}}_n^-\equiv \mtau \left[\frac{1}{n} (\mu_1\mtau)^{n-1} -1\right].
\label{eq:minmaxbounds}
\end{align}
Notably, in dimensionless units of $1/\mu_1$, these bounds are
functions of $\mu_1\mtau$ only. In other words, the mean $\mtau$ sharply bounds the expected range of extreme
first-passage times even when it is \emph{a priori} not a representative statistic 
due to the presence of multiple time-scales. 
Therefore, these results motivate an estimation of $\mtau$ via $\ntau$ even in the presence
of many time-scales.
The validity and sharpness 
is depicted in Fig.~\ref{Fig5}b,c for expected deviations of the maximum (blue)
and the minimum (orange) for the Markov jump network from Fig.~\ref{Fig1}c.
Each data point corresponds to one particular realization of the 
jump process (see Sec.~\ref{sec:markov_network}) where 
transition rates are drawn randomly under the constraint of detailed balance. 
\blue{Note that the ``squeeze'' bounds \eqref{eq:minmaxbounds}
can be saturated (for a discussion see \cite{bebon2023controlling})}.

Considering that the maximal and minimal first-passage time
in a sample of $n$ i.i.d.~realizations can be viewed as representing
the fastest and slowest first-passage times among the $n$
independent searchers, respectively, these results provide insight into
the slowest and fastest first-passage events of $n$ independent searchers.
Therefore, it is anticipated that these bounds will prove to be relevant
in target-searches with multiple searches and in the few-encounter limit.

\section{Discussion}
\label{sec:conclusions}
To robustly asses the uncertainty of inferred 
empirical first-passage times
in the small-sample regime in a general setting,
one needs to go beyond conventional approaches relying on asymptotic reasoning
or require prior belief.
In this chapter, we presented one approach that achieves reliable error estimates
by tackling the issue from a non-asymptotic concentration-of-measure perspective 
that builds on spectral properties of the first-passage problem.

In particular, we put forward a framework that
allows to bound uncertainties associated with empirical first-passage times
in terms of deviation probabilities that hold regardless of sample size.
Subsequently, we showcased how these bounds can be applied by constructing 
non-asymptotic confidence intervals, which allowed us to tackle
the critical question of what sample size is required to ensure
tolerable errors.

In the last part, we addressed the challenging situation
of many relevant time-scales, where it is no longer 
adequate to rely solely on the mean first-passage time for an appropriate description.
In doing so, we presented sharp non-asymptotic bounds on average minimal and maximal deviations 
around the mean, and argued that these results have direct implications
for target-search processes with multiple searchers.

To conclude, the presented framework of non-asymptotic uncertainty quantification 
is applicable to the broad class of reversible
Markov dynamics that include discrete Markovian jump-processes in any
dimension and Markovian diffusion in effectively one-dimensional potential
landscapes. Accordingly, it will be interesting to see how the developed theory
can be turned into new statistical tests in the future. Moreover, 
extensions to, for example, irreversible (i.e., driven) dynamics 
and more complicated observables will be carried out in the future.

\begin{acknowledgement}
Financial support from Studienstiftung des Deutschen Volkes (to R.~B.) and the German
Research Foundation (DFG) through the Emmy Noether
Program GO 2762/1-2 (to A.~G.) is gratefully acknowledged.
\end{acknowledgement}

\end{document}